\documentclass[lettersize,journal]{IEEEtran}
\usepackage{amsmath,amsfonts}
\usepackage{algorithmic}
\usepackage{algorithm}
\usepackage{array}
\usepackage[caption=false,font=normalsize,labelfont=sf,textfont=sf]{subfig}
\usepackage{textcomp}
\usepackage{stfloats}
\usepackage{url}
\usepackage{bm}
\usepackage{verbatim}
\usepackage{makecell}
\usepackage{hyperref}
\usepackage{graphicx}
\usepackage{multirow}
\usepackage{booktabs}
\usepackage{amsmath}
\usepackage{cite}
\usepackage{arydshln}
\usepackage{xcolor}
\begin{document}

\title{PhiNet: Speaker Verification with Phonetic Interpretability}






\author{Yi Ma, \IEEEmembership{Student Member, IEEE}, Shuai Wang*, \IEEEmembership{Senior Member, IEEE}, Tianchi Liu, \IEEEmembership{Student Member, IEEE},
Haizhou Li, \IEEEmembership{Fellow, IEEE}

\thanks{This work was supported in part by National Natural Science Foundation of China (Grant No. 62401377 and No. 62271432); 
Shenzhen Science and Technology Program (Shenzhen Key Laboratory, Grant No. ZDSYS20230626091302006); 
Program for Guangdong Introducing Innovative and Entrepreneurial Teams (Grant No. 2023ZT10X044). (*Corresponding author: Shuai Wang)

Yi Ma and Tianchi Liu are with the Department of
Electrical and Computer Engineering, National University of Singapore,
119077, Singapore. (e-mail: mayi@u.nus.edu and tianch\_liu@u.nus.edu) 

Tianchi Liu is also with LIGHTSPEED, 068897, Singapore. 

Shuai Wang is with School of Intelligence Science and Technology, Nanjing University, Suzhou, 215163, China; also with Shenzhen Loop Area Institute, Shenzhen, China. (e-mail: shuaiwang@nju.edu.cn)

Haizhou Li is with School of Artificial Intelligence, Shenzhen Research Institute of Big Data, The Chinese University of Hong Kong, Shenzhen, Guangdong 518172,
China; also with Shenzhen Loop Area Institute, Shenzhen, China(haizhouli@cuhk.edu.cn)

The code can be found in: https://github.com/mmmmayi/PhiNet

}}

\markboth{Journal of \LaTeX\ Class Files,~Vol.~14, No.~8, August~2021}%
{Shell \MakeLowercase{\textit{et al.}}: A Sample Article Using IEEEtran.cls for IEEE Journals}


\maketitle

\begin{abstract}

Despite remarkable progress, automatic speaker verification (ASV) systems typically lack the transparency  required for high-accountability applications. Motivated by how human experts perform forensic speaker comparison (FSC), we propose a speaker verification network with phonetic interpretability, \textit{PhiNet}, designed to enhance both local and global interpretability by leveraging phonetic evidence in decision-making.
For users, PhiNet provides detailed phonetic-level comparisons that enable manual inspection of speaker-specific features and facilitate a more critical evaluation of verification outcomes.
For developers, it offers explicit reasoning behind verification decisions, simplifying error tracing and informing hyperparameter selection.
In our experiments, we demonstrate PhiNet’s interpretability with practical examples, including its application in analyzing the impact of different hyperparameters.
We conduct both qualitative and quantitative evaluations of the proposed interpretability methods and assess speaker verification performance across multiple benchmark datasets, including VoxCeleb, SITW, and LibriSpeech.
Results show that PhiNet achieves performance comparable to traditional black-box ASV models while offering meaningful, interpretable explanations for its decisions, bridging the gap between ASV and forensic analysis.

\end{abstract}


\section{Introduction}

Automatic speaker verification (ASV) seeks to validate an identity claim based on the speaker's voice. Successful ASV models include Gaussian Mixture Model-Universal Background Model (GMM-UBM)~\cite{auckenthaler1999,kenny2003new,parris1994,kajarekar2001}, i-vector\cite{5545402}, and recently neural network-based approaches such as x-vector\cite{snyder2018x}, r-vector\cite{zeinali2019but} and ECAPA-TDNN~\cite{desplanques2020ecapa}, which have dramatically improved both accuracy and robustness in speaker verification tasks. 

While forensic speaker comparison (FSC) also seeks to validate identity claims, it emphasizes collecting evidence to support decision-making~\cite{foulkes2012forensic,nolan1983phonetic,morrison2019introduction}
. Unlike ASV, FSC is inherently complex, with decisions often subject to debate. Speech scientists acknowledge that voice alone cannot be used to confirm identity with absolute certainty~\cite{foulkes2012forensic}. Consequently, forensic experts rely on detailed and persuasive evidence to support their conclusions, particularly in legal contexts, where findings must be presented in a manner understandable to judges and open to scrutiny.
Despite their shared objective, ASV and FSC approach speaker verification differently. FSC employs scientific tools such as voice activity detectors, spectrogram analyzers, and pitch trackers to form expert opinions, whereas ASV systems typically function as black-box models~\cite{6004980,tian2017,nagavi2023voice,morrison2019introduction}. These systems typically produce a single similarity score to indicate a match between voice samples, offering minimal  interpretability~\cite{wang23ha_interspeech}. This lack of transparency undermines the trust of the user due to concerns about the reliability of the system and potential biases. 



Providing evidence for speaker verification not only builds user trust but also facilitates error tracing and correction through explicit reasoning.
In contrast, the opaque nature of ASV prevents model designers from identifying the root causes of errors, limiting opportunities for direct improvements~\cite{arrieta2020explainable}.
To address these limitations of black-box ASV, we aim to bridge the gap between ASV and FSC, making ASV results useful as forensic evidence, where transparency, reliability, and accountability are crucial.

A key distinction between ASV and FSC lies in how FSC extracts evidence. 
FSC extracts evidence primarily through three methods. First, visual spectrogram inspection, based on the hypothesis that each voice has unique fingerprint-like features~\cite{KERSTA}, which is now regarded as scientifically unreliable due to its subjectivity~\cite{bolt1973speaker,hollien2013acoustics}. Second, linguistic-based analysis, widely established in forensic settings~\cite{foulkes2012forensic,rose2002forensic,hollien2013acoustics}, which decomposes speech into units such as consonants, vowels, and intonation for acoustic and auditory analysis. Third, automatic techniques increasingly integrated with linguistic analysis, which quantify features like fundamental frequency (f0) and formants, and expressing results as likelihood ratios~\cite{hirson2012speech,hudson2007f0,braun1995fundamental}. Across all FSC methods, phonetic information remains crucial for producing reliable and defensible verification evidence. Building on this insight, the concept of phone-by-phone comparison has been previously investigated in \cite{yang2022phone} to refine speaker representations.

Motivated by the evidential approach in FSC, we introduce the \emph{Speaker Verification with Phonetic Interpretability Network} (PhiNet), which enhances ASV by providing linguistic-based evidence alongside the verification score.
On one hand, this evidence deconstructs the final score into observable phonetic-level comparisons, highlighting the contribution of each segment. We refer to this as \textit{local interpretability} since it operates at the trial level.
On the other hand, PhiNet offers \textit{global interpretability} by ranking phonemes based on their distinctiveness for speaker identification, thereby assisting researchers in understanding potential system biases.
Notably, our research reveals that the distinctiveness rank of phonemes in our model aligns with their importance observed in FSC. For example, our model consistently identifies nasals and vowels as crucial for speaker verification, while also capturing significant variability in the distinctiveness of fricative sounds. In summary, this paper makes the following contributions:
\begin{itemize}


\item To bridge the gap between ASV and FSC, especially the black-box nature of current ASV systems, we propose PhiNet, the first self-interpretable speaker verification network that explains its decision-making process.
\item PhiNet leverages phoneme distinctiveness, offering dual interpretability: (a) Locally, PhiNet reveals each phoneme's contribution to the verification decision of an for individual trial; (b) Globally, it identifies patterns consistent with FSC heuristics, demonstrating which specific phoneme and its pronunciation patterns significantly contribute to the verification outcome.
\item Unlike most ASV models that only focus on making verification decision, PhiNet implements a training scheme that simulates the verification process, ensuring consistency between training and verification phases and enhancing the reliability of interpretable decisions.
\item Through comprehensive experimental analysis, we uncover insights that explain previously observed phenomena in speaker recognition, such as the poor robustness of short-duration speaker recognition and the impact of training segment length on verification performance.
 
\end{itemize}
The remainder of the paper is organized as follows. In Section~\ref{works} we introduce the related work. In Section~\ref{proposed} we formulate PhiNet and its training scheme.   Section~\ref{setting} discusses the experimental setup. In Section~\ref{result}, we report the experimental results. Section~\ref{conclusion} concludes the study.

\section{Related work}
\label{works}
We first introduce the concept of neural network interpretability in the general field of machine learning, then discuss the current research on neural network interpretability for speaker-related tasks.
\subsection{Neural Network Interpretability}
Interpretability in neural networks, also known as explanability, has gained prominence in deep learning due to the black-box nature of these models. It refers to methods that reveal how a model reaches its conclusions~\cite{zhang2021survey,arrieta2020explainable,saeed2023explainable}. In this section, following the taxonomy in~\cite{zhang2021survey}, we classify interpretability methods along two axes of passive vs. active and global vs. local.

\subsubsection{Passive vs. Active}
Passive interpretation methods present explanations from a model’s existing predictions without altering the neural network's architecture or training process. In contrast, active methods, or self-interpretable techniques, modify the model architecture or training process for interpretability, which are considered more accurate and direct~\cite{wang2021self,alvarez2018towards}.
\paragraph{Passive Interpretation}
In early stages, extracting logical rules or building decision trees were the approach making model decisions more interpretable~\cite{craven1995extracting,andrews1995survey}. 
Other popular methods include feature attribution, divided into backpropagation-based (e.g., gradients to assess feature impact~\cite{simonyan2013deep,selvaraju2017grad,sundararajan2017axiomatic,wang2020score,9462463}) and model-agnostic approaches (e.g., LIME~\cite{ribeiro2016should} and Shapley Values~\cite{shapley1953value,lundberg2017unified}, which use perturbation-based analysis~\cite{petsiuk2018rise,fong2019understanding}).
Techniques like activation maximization reveal high-level features by optimizing inputs to maximize neuron activations~\cite{wang2018visualizing,zhang2018interpretable,fong2018net2vec,dalvi2019one}. 
Additionally, data-centric methods explain predictions by identifying influential training samples~\cite{yeh2018representer,koh2017understanding}.

\paragraph{Active Interpretation} Prototype-based methods, such as ProtoPNet~\cite{li2018deep,chen2019looks,rymarczyk2022interpretable,alvarez2018towards} and its variants (e.g. Deformable ProtoPNet~\cite{donnelly2022deformable}, PIP-Net~\cite{nauta2023pip}  and XProtoNet~\cite{kim2021xprotonet}), compare inputs to learned prototypes, producing intuitive explanations. 
Based on this, to ensure explanations remain stable under transformations of the input data, SITE~\cite{wang2021self}  integrates transformation-equivariant prototypes into the model’s architecture.
Concept Bottleneck Models~\cite{koh2020concept} use a method 
similar to prototype, and the language models advances this to generate bottleneck concepts without human-labeled attributes~\cite{yang2023language}. 
Feature ranking~\cite{wojtas2020feature,weinberger2020learning} and component modeling~\cite{wang2022interpretability,nanda2023progress,shah2024decomposing} further enhance interpretability by prioritizing relevant features and analyzing internal components.

\subsubsection{Local vs. Global}
At the same time, interpretable neural networks can be categorized into local and global forms based on the scope of their interpretability. Local interpretation aims to provide insights into why the model produced a particular output for a specific input. In contrast, global interpretability focuses on understanding the overall decision logic of a model, which is consistent for all samples. However the topic of global interpretability remains relatively underexplored.
\paragraph{Local Interpretation}
Techniques in this category often rely on attribution methods, which highlight the importance of individual input features. For example, one popular local interpretation method is the gradient-based~\cite{selvaraju2017grad,sundararajan2017axiomatic,9462463} and perturbation-based~\cite{shapley1953value,lundberg2017unified,ribeiro2016should} attribution approaches we mentioned above, such as Grad-CAM~\cite{selvaraju2017grad} and LIME~\cite{ribeiro2016should} that identify the regions of an input that most influenced a prediction. Rule-based methods, such as those in~\cite{ribeiro2018anchors} and~\cite{dhurandhar2018explanations}, also provide local interpretability. For example,~\cite{ribeiro2018anchors} introduces anchors, the high-precision, model-agnostic if-then rules that ensure a prediction remains consistent when their conditions are met. Similarly,~\cite{dhurandhar2018explanations} proposes a contrastive explanation framework that identifies minimal features both necessary to justify a prediction and those whose absence is critical to preserving it.
\paragraph{Global Interpretation}
 Zhang et al.~\cite{zhang2018interpretable} develops a method to transform traditional CNNs into interpretable CNNs, ensuring that each high-level filter explicitly represents distinct object parts. Wojtas and Cheny~\cite{wojtas2020feature} integrated feature subset selection with model performance optimization, providing global insights into feature contributions and enhancing explainability through a stochastic optimization strategy. Pedapati et al.~\cite{pedapati2020learning} introduced a technique for constructing globally interpretable models that closely align with local contrastive explanations derived from black-box models. Fong and Vedaldi~\cite{fong2018net2vec} proposed the Net2Vec framework, which maps semantic concepts to vector embeddings aligned with neural network filter activations. Unlike local interpretability, global interpretability uncovers patterns and rules that define a network's decision-making across the entire dataset, offering a comprehensive, systemic understanding of its behavior. Furthermore, it transcends individual predictions by revealing the fundamental principles learned by the network~\cite{zhang2021survey}.

\subsection{ Interpretability for Speaker-Related Tasks}
The topic of interpretability has recently gained attention in speaker-related tasks as well. Most existing studies have primarily focused on explaining closed-set speaker recognition tasks, while open-set speaker verification tasks remain largely unexplored. 
For speaker recognition tasks, identifying the time-frequency bins within input features that are most critical to the network has become a popular approach for interpretability. For example, Li et al.~\cite{li2022reliable} investigated the reliability of three Class Activation Map-based algorithms as visualization tools for deep speaker recognition models. Building on this, they utilized these methods to analyze the effects of data augmentation on model robustness~\cite{li2023visualizing} and to evaluate the contributions of various phoneme categories to deep speaker models~\cite{li2024phonemes}. 
Similarly, Zhang et al.~\cite{zhang2023study} also used visualization as an interpretability tool to analyze speaker recognition models, using five attribution methods, including Integrated Gradients~\cite{sundararajan2017axiomatic} and SHAP~\cite{lundberg2017unified}, to visualize voiceprint features within the ECAPA-TDNN framework. 
Similarly, Thebaud et al.~\cite{thebaud2024phonetic} leveraged Integrated Gradients to assess the contribution of various phoneme categories to the performance of speaker verification models. Beyond visualization,  Ben-Amor et al.~\cite{ben2023describing} proposed a novel explainability framework using the Binary-Attribute Likelihood Ratio (BA-LR), which decomposes speaker recognition scoring into interpretable sub-processes based on binary speech attributes.

The work most closely related to the method we pursue~\cite{ma2025expo} is the explainable attribute-based speaker verification framework proposed by Wu et al.~\cite{wu2024explainable}. In our previous work~\cite{ma2025expo}, we introduced phonetic traits to represent the speaker-specific information embedded in each phoneme. However, these traits were utilized passively to explain the speaker network rather than being directly involved in the prediction process. This approach may lead to a gap in faithfulness between the provided explanations and the actual working mechanism. In~\cite{wu2024explainable}, the authors adopt a two-stage pipeline leveraging Concept Bottleneck Models~\cite{koh2020concept} to integrate explainability directly into the speaker verification process. However, their system's performance is limited by a small set of attributes that lack generalization, resulting in a significant {compromise} of accuracy. 

It is worth noting that most existing interpretability studies for speaker-related tasks adopt passive and local interpretability methods. While these methods are effective in explaining specific predictions, they often fail to provide insights into the overall decision-making logic of the model. Additionally, while visualization tools have proven effective for analyzing closed-set tasks, their applicability remains limited when it comes to open-set tasks like speaker verification, where the ability to generalize to unseen speakers is critical.



\section{Methodology}
\label{proposed}

\begin{figure*}[htbp]
\centering
\includegraphics[width=16cm]{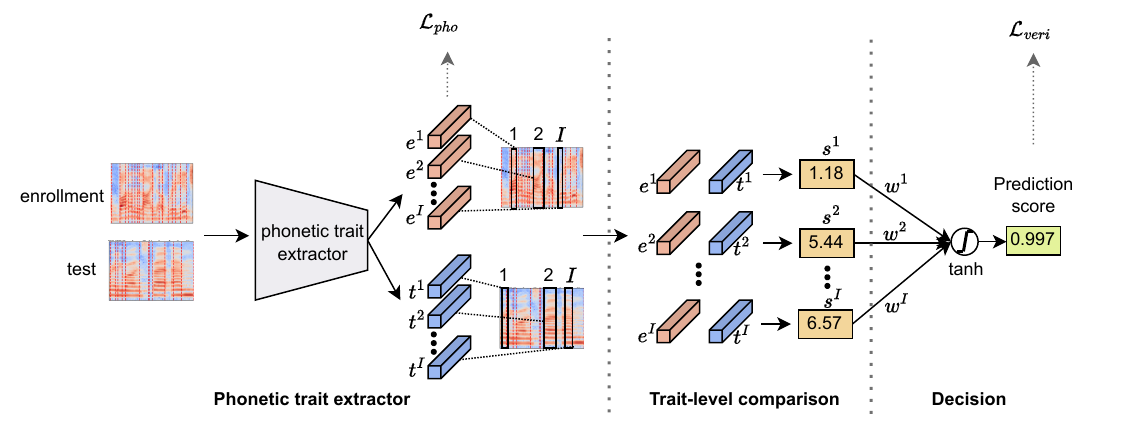}
\caption{Block diagram of the proposed framework for speaker verification with phonetic interpretability.}
\label{diag}
\end{figure*}
We introduce a novel interpretability framework designed to provide both global and local insights into the decision-making process of speaker verification.
Fig.~\ref{diag} shows the overall architecture of our proposed PhiNet, which is composed of three key components: the phonetic trait extractor, trait-level comparison layer, and decision layer. 

\subsection{Verification with Phonetic Interpretability}
\label{process}

Our PhiNet performs speaker verification by comparing the enrollment and test utterance to generate a score that quantifies the similarity between the speakers of these two utterances.  Let's assume a phonetic inventory of  $I$ phonemes.  For a pair of utterances, the phonetic trait extractor generates a set of $I$ phonetic traits, denoted as $\bm{E} = \{\bm{e}^i\}^I_{i=1}$ and $\bm{T} = \{\bm{t}^i\}^I_{i=1}$ {for the pair}. Each $\bm{e}^i$ and $\bm{t}^i$ captures the speaker-specific information associated with phoneme $i$ in an utterance. 

In the phonetic trait comparison layer, we compute the phonetic score $s^i$ for each phoneme $i$ based on the cosine similarity between $\boldsymbol{e}^i$ and $\boldsymbol{t}^i$. This score quantifies the similarity of speaker information for phoneme $i$ across the two utterances.

Considering that the ability of  phonemes varies in terms of speaker characterization, higher weights should be assigned to phonemes that are more discriminative. In this way, phonetic score for these high-weight phonemes dominates the final score. Conversely, phonemes that contribute less to speaker identification are assigned lower weights, thereby reducing their influence on the final score. The weight, quantified and denoted as $w^i$, is jointly learned with our network. The final score for these two utterances  is computed as the weighted average of $s^i$, where the weights are given by the $w^i$.

This process interprets verification as a comparison of speaker information within each phonetic segment, 
allowing users to examine the spectrograms of specific phonetic segments as evidence.
Additionally, $w^i$, serves as a global interpretability factor, which reflects the potential cognitive bias of the model. 
The phonetic weight allows users to understand the model’s reliance on each phoneme, providing transparency to check if the model works reasonably.

\subsection{Architecture}
\label{module}

\begin{figure}[htbp]
\centering
\includegraphics[width=8.5cm]{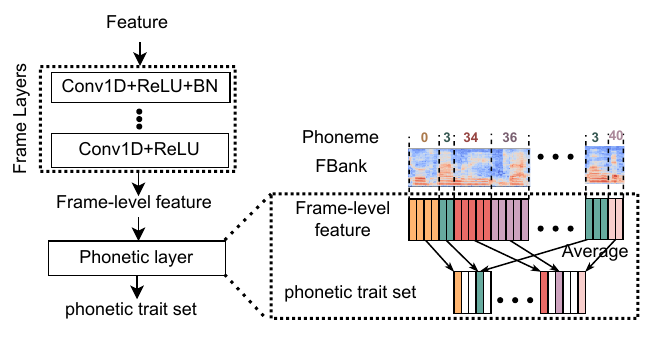}
\caption{Block diagram of the phonetic trait extractor.}
\label{extractor}
\end{figure}
\subsubsection{Phonetic trait extractor}
As shown in Fig.~\ref{extractor},  the phonetic trait extractor consists of two main components: frame-level layers and phonetic-level layers, creating a fine-to-coarse hierarchical structure. Given that frame-level feature extractors are commonly used in  speaker verification networks, without loss of generality, we adopt the frame-level layers in ECAPA-TDNN to generate $D$-dimensional frame-level feature. Aligned with phonetic boundaries provided by a pre-trained  phoneme recognizer in textless mode~\cite{zhu2022charsiu}, frame-level features  are then passed to the phonetic-level layer. Using these phonetic boundaries, we calculate the average of frame-level features within phoneme $i$ segment.  This averaged representation is treated as the phonetic trait, which we consider to capture the speaker-specific characteristics associated with phoneme $i$. For phonemes that do not appear in a given utterance, the associated phonetic trait is set to $\mathbf{0}$ for the absent phoneme $i$. This design is consistent with our prior work~\cite{ma2025expo}, we invite readers to consult the full paper and accompanying code\footnote{\url{https://github.com/mmmmayi/ExPO}} for a deeper understanding of the implementation.

\subsubsection{Trait level comparison}

\begin{figure}[htbp]
\centering
\includegraphics[width=8.5cm]{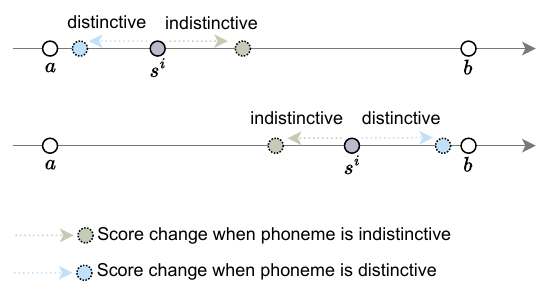}
\caption{Illustration of two conditions of adjusting phonetic scores based on phoneme distinctiveness. The top figure shows the adjustment when the phonetic score is closer to $a$, while the bottom figure depicts the adjustment when the score is closer to $b$. Here, $a$ and $b$ represent the scores for completely different speakers and the same speaker, respectively.} 
\label{adjust}
\end{figure}
We compute the cosine similarity of each pair of $(\bm{e}^i, \bm{t}^i)$.
Due to adapting the frame-level layers of ECAPA-TDNN, the extracted phonetic traits are constrained to be nonnegative, making the resulting cosine similarity fall within the range $[0,1]$. A transformation function is applied on the cosine similarity, which spans the result into range of $[-\infty, +\infty]$:
\begin{gather}
d^i=\frac{\bm{e}^i \cdot \bm{t}^i}{\|\bm{e}^i\| \|\bm{t}^i\|},\label{map}\\
s^i=f_2\left(\phi\left(f_1(d^i)\right)\right)\cdot d^i,\label{map2}
\end{gather}
where $f_1$ and $f_2$ are linear layers with an intermediate dimension $c$ and $f_2$ has no bias term. 
Specifically, 
$f_{1}: \mathbb{R}^{1} \rightarrow \mathbb{R}^{c}$, 
and $f_{2}: \mathbb{R}^{c} \rightarrow \mathbb{R}^{1}$,
where $c$ controls the dimension of the mapping vector.
Both $f_1$ and $f_2$ are shared across all phoneme.  $\phi$ is the Tanh activation function.
\subsubsection{Decision layer}
In the decision layer, we generate the final score using $[s^1, s^2, ... ,s^I]$ and $[w^1, w^2, ... ,w^I]$. The weight $w^i$ is designed to amplify the contribution of $s^i$ if phoneme $i$ is distinctive for speaker verification, otherwise to reduce the contribution of $s^i$. This effect is illustrated in Fig.~\ref{adjust}. Consider an example where  $a$ represents the similarity score for completely different speakers and  $b$ represents the score for identical speakers, with the constrains $a<b$ and $s^i\in [a,b]$. We assume that the midpoint between $a$ and $b$ serves as the threshold. If $s^i$ is closer to $a$, the trial is classified as involving different speakers. Conversely, it is considered to originate from the same speaker.
When $s^i$ is closer to $a$ before adjustment and phoneme $i$ is distinctive, $w^i$  should adjust $s^i$ shift further toward to $a$. Conversely, for a non-distinctive phoneme $i$, The weighted $s^i$ should have a larger distance to $a$ than before. It's worth to mention that even for a non-distinctive phoneme  $i$, the adjustment should not contradict the conclusion derived from the original $s^i$. In other words, if $|s^i-a|\leq|s^i-b|$ before applying $w^i$, this relationship should hold true after adjustment.

A straightforward approach to implement this functionality is to ensure $s^i \in \mathbb{R}$ and $ w^i \in [0, 1]$, while assuming a threshold of 0 for distinguishing between target and non-target trials.
After applying min-max normalization to a learnable vector $\bm{\hat{w}}$ of size $I\times1$, we obtain $\bm{w}$, which serves as the phonetic weights:
\begin{equation}
    \label{weight}
    \bm{w}=\frac{\bm{\hat{w}}-\min(\bm{\hat{w}})}{\max(\bm{\hat{w}})-
    \min(\bm{\hat{w}})+\epsilon},
\end{equation}
where $\epsilon$ represents an arbitrarily small positive number.

The element of $\bm{w}$, denoted as $w^i$, is used in the computation of the prediction score $y$:  
\begin{equation}
\label{y}
    y=\phi \left(\frac{1}{\sum_i^I \left[\mathbb{I}\left(\bm{e}^i \neq \mathbf{0} \land \bm{t}^i \neq \mathbf{0}\right)\cdot w^i\right]}\sum^I_{i=1} w^i s^i\right),
\end{equation}
where $\land$ represents the logical AND operator.  \(\mathbb{I}(\cdot)\) is the indicator function, having 1 if the condition is true and 0 otherwise.
$\bm{w}$ is optimized jointly with the network during training, allowing the model to learn the relative distinctiveness of each phoneme for accurate speaker verification.
\subsection{Training Scheme}
\label{training}
We have discussed  the process above where only one pair of utterances are compared. In this section, we describe the training scheme with $K$ pair of utterances fed into the model as a batch. Specifically, we randomly selected $K$ speakers, and each speaker contributes two randomly selected utterances. We viewed these two utterances as enrollment and test utterance respectively to simulate the inference process. 

Given a batch of utterances, the phonetic trait for each speaker is denoted as $\bm{E}_k=\{\bm{e}^i_k\}^I_{i=1}$ and $\bm{T}_k=\{\bm{t}^i_k\}^I_{i=1}$ respectively, with $k=1,2,\dots,K$. The score for enrollment utterance from the speaker $k$ and test utterance from the speaker $j$ is denoted as $y_{kj}$.

In this way, the network is expected to derive a high value for $y_{kj}$ if $k=j$, otherwise, low value for $y_{kj}$. Based on this hypothesis, we propose the verification loss $\mathcal{L}_{veri}$:
\begin{equation}
\label{veri}
\mathcal{L}_{veri}=- \frac{1}{K} \sum_{k=1}^K  \log \frac{\exp (y_{kk})}{\sum_{j=1}^K \exp(y_{kj})}.
\end{equation}
The $\mathcal{L}_{veri}$ is applied on each enrollment utterance in this batch and the model is forced to identify the test utterance from the same speaker.

We also propose the phonetic trait loss $\mathcal{L}_{pho}$ to strengthen the trait consistency for the same speaker while discriminates the traits between different speakers: \begin{equation}
\label{pho}
\begin{split}
\mathcal{L}_{pho}&= \frac{\alpha }{N_1} \sum_{i=1}^I \sum_{k=1}^{K} \mathbb{I}( \bm{e}_k^i \neq \mathbf{0} \land \bm{t}_k^i \neq \mathbf{0}) || \bm{e}_k^i - \bm{t}_k^i ||^2 \\
&-\frac{\beta}{N_2} \sum_{i=1}^I \sum_{k=1}^{K}  \min_{\substack{ j \neq k}} \mathbb{I}(\bm{e}_k^i \neq \mathbf{0} \land \bm{t}_j^i \neq \mathbf{0}) || \bm{e}_k^i - \bm{t}_j^i ||^2,
\end{split}
\end{equation}\\[0.8em]
\(N_1 = \sum_{i=1}^I \sum_{k=1}^{K} \mathbb{I}(\bm{e}_k^i \neq \mathbf{0} \land \bm{t}_k^i \neq \mathbf{0})\) and  \(N_2 = \sum_{i=1}^I \sum_{k=1}^{K} \\[0.8em] \mathbb{I}(\bm{e}_k^i \neq \mathbf{0} \land \bm{t}_j^i \neq \mathbf{0})\) are two denominators. $\alpha$ and $\beta$ are hyper-parameters to balance the loss for inter-speaker and intra-speaker trait. 
The first term of Eq. (\ref{pho}) enforces the consistency of phonetic traits between matched pairs ($\bm{e}^i_k$, $\bm{t}^i_k$) for the same phoneme,  while the second term maximizes the dissimilarity between $\bm{e}^i_k$ and its most similar unmatched counterpart $\bm{t}^i_k$, retaining only the smallest distance among unmatched pairs.

The model is optimized  by $\mathcal{L}_{veri}$ and $\mathcal{L}_{pho}$ simultaneously, and the total loss is defined as
\begin{equation}
\mathcal{L}_{all}=\gamma\mathcal{L}_{veri}+\mathcal{L}_{pho},
\end{equation}
where $\gamma$ is the balancing weight for two loss values.
\section{Experiments}
\label{setting}
\subsection{Dataset}
Our experiments are carried out on the VoxCeleb 1~\cite{nagrani2017voxceleb}, VoxCeleb 2~\cite{chung2018voxceleb2}, the Speaker In the Wild (SITW)~\cite{mclaren16_interspeech}, and LibriSpeech~\cite{7178964}.
\subsubsection{VoxCeleb 1}
VoxCeleb 1 is a large-scale audio-visual dataset widely used for speaker recognition and verification tasks. It contains 153,516 utterances collected from 1,251 speakers, extracted from YouTube videos. VoxCeleb 1 is collected under real-world conditions and includes variations in age, accent, gender, and language. The dataset is divided into training set and test set, and three trial lists, including Vox1-O, Vox1-E and Vox1-H, are provided. The training set consists of audio data from 1,211 speakers, containing more than 148,642 utterances, and the test set contains 4,874 utterances from 40 speakers. The trial list Vox1-O is generated from the test set, while Vox1-E and Vox1-H are drawn from the training set. We use the training set of VoxCeleb1 to fine-tune the hyper-parameters  and Vox1-O  to evaluate the system in ablation experiments.   

\subsubsection{VoxCeleb 2}
VoxCeleb 2 is an extension of the VoxCeleb 1 dataset, providing a larger and more diverse collection of audio-visual data for speaker recognition research. It contains over 1 million utterances from 6,112 speakers. We name the training set of VoxCeleb2 as ``Vox2'' in following sections, which consists of 5,994 speakers and 1,092,009 utterances. 

\subsubsection{SITW}
Similarly to VoxCeleb 1 and 2, SITW is a large, publicly available dataset designed to reflect real-world conditions in speaker recognition research. The SITW dataset contains audio recordings of 299 speakers, with approximately 6,000 utterances collected from open-source media~\cite{mclaren16_interspeech}. To evaluate our system, we use the the official trial lists of ``core-core" condition from the evaluation  parts of the dataset (SITW-eval), which contain 721,788 trials, respectively. Unlike the VoxCeleb1-O/E/H test sets, which have a balanced number of target and non-target trials, SITW exhibits a significantly imbalanced distribution. Specifically, the counts for target and non-target trials in SITW-eval are 3,658 and 718,130 respectively. 

\subsubsection{LibriSpeech}
The LibriSpeech is a widely used corpus of read English speech.  It consists of approximately 1,000 hours of speech data derived from audiobooks in the public domain. The dataset includes recordings from a wide range of speakers, covering various ages, accents, and genders. LibriSpeech also provides text transcriptions for all the recordings. Our phoneme recognizer is pre-trained on the LibriSpeech, with the phonetic inventory containing 39 units
in the CMU phoneme set~\cite{phoset} and a non-verbal label [N-V]. Therefore, $I = 40$ in our paper.

We also use the LibriSpeech as one of our test set to evaluate the generalization ability of our model.  Since there is no official trial list for LibriSpeech, we randomly generated one target and one non-target trial for each utterance in the train-clean-100, train-clean-360, dev-clean, and test-clean subsets, resulting in a total of 275,752 trials.

\subsection{Experimental Setup}
Our system is implemented using WeSpeaker toolkit~\cite{wang2022wespeaker}. The feature is an 80-dimensional log-mel filterbank (FBank) with a window size of 400 and a hop length of 160.  The frequency range is from 20 Hz to 7600 Hz. 
We use ECAPA-TDNN~\cite{desplanques2020ecapa} as the backbone and follow the configuration of ``ecapa-tdnn" in WeSpeaker to train our model. 
Following the setting of ECAPA-TDNN, $D$ is set as 1,536.  
We increase the number of epochs to 13,650 so that the number of samples used for training is approximately equal to the toolkit settings.  Our model is trained with Stochastic Gradient Descent (SGD) as the optimizer.  The initial learning rate is set to 0.1 and exponentially decays to 0.00005. To ensure the learning rate aligns with the baseline configuration, $\gamma$ is set to 0.5. The MUSAN dataset (babble, noise, music)~\cite{snyder2015musan} and the RIR dataset (reverb)~\cite{ko2017study} are used for augmentation. The samples are subjected to 0.6 probability of not being augmented, 0.2 probability of adding noise, and 0.2 probability of adding reverberation during training. The duration of each training sample is set to 3 seconds, with $K$ and $c$ assigned values of 128 and 2, respectively, unless stated otherwise.

\subsection{Baseline Setting}
The ECAPA-TDNN serves as the black-box model for performance comparison. For the baseline setup, we first adopt the pre-trained ECAPA-TDNN released in the open-source project wespeaker\footnote{\url{https://wenet.org.cn/downloads?models=wespeaker&version=voxceleb_ECAPA512.zip}}, which matches the same model configuration as the one used in our experiments.

To further eliminate the possible confounding effects brought by different loss functions and pooling mechanisms, we additionally reproduced an ECAPA-TDNN model trained following exactly the same training pipeline as ours.
Specifically, to ensure a fair comparison between the ECAPA-TDNN and our proposed model, we adopted a metric learning approach for training ECAPA-TDNN, specifically utilizing the Angular Prototypical Loss function~\cite{wang2017deep}. 
In addition, there are two further differences between the reproduced ECAPA-TDNN and the released pre-trained model. To avoid introducing phoneme distortion and ensure that augmentation does not become an additional confounding variable, we excluded speed perturbation from the reproduced model. Furthermore, we employed temporal statistics pooling instead of the attentive statistics pooling used in the pre-trained model, so as to minimize the contribution of attention-based pooling biases. These adjustments allow the comparison to be primarily focused on the difference between an interpretable system and a black-box system, rather than differences arising from augmentation choices or pooling mechanisms.
To facilitate reproducibility, we release our code at \href{https://github.com/mmmmayi/LightWespeaker\_Prototypical}{https://github.com/mmmmayi/LightWespeaker\_Prototypical}.

\subsection{Evaluation Metric}
\label{eval metric}

\subsubsection{Evaluating speaker verification performance}
We report the speaker verification performance in term of equal error rate (EER) and minimum detection cost function (minDCF) with $p_{target}=0.05$.

\subsubsection{Evaluating interpretability}
Our approach of global interpretability involves ranking the individual contributions of phonetic traits to the final prediction score. 
To verify the reliability of this ranking, we perform the {leave-$i$th-phoneme-out} experiments on the input spectrogram.  Specifically, following this ranking, we sequentially remove the segments within each phoneme from the input spectrogram. 
The hypothesis is that if a phoneme significantly influences prediction, its removal will result in a substantial decrease in performance, whereas the removal of phonemes with lower rankings will yield negligible changes. This performance variation due to segment removal serves as a qualitative validation of the system’s global interpretability.

Local interpretability relies on the hypothesis that each phonetic trait encodes speaker information specific within its corresponding phonetic segments.
To evaluate the faithfulness of this hypothesis, we employ a similar evaluation method to that used for global interpretability. 
Instead of removing specific phonetic segments from the spectrogram, we exclude the corresponding phonetic traits during the decision process. If a phonetic trait truly represents speaker information localized to specific segments, the performance impact of removing the trait should closely align with that of removing the corresponding spectrogram segments.
To quantify this faithfulness, we calculate the mean absolute difference between the performance changes caused by these two types of removal across all phonemes, defining as the \textit{fidelity score}.  A lower fidelity score reflects better alignment, indicating that phonetic traits accurately represent localized speaker information.

 \section{Results}
\label{result}
\subsection{Local Interpretation}
\begin{figure*}[htbp]
\centering
\includegraphics[width=14cm]{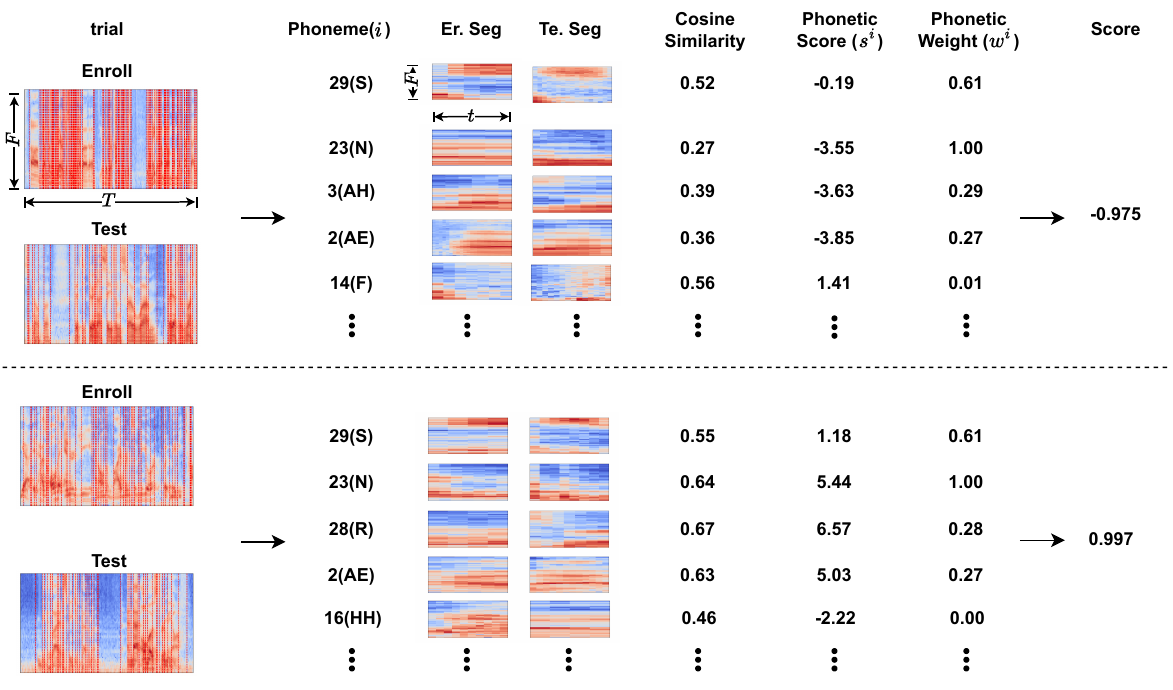}
\caption{Visualization of the decision making process for a non-target trial (top) and a target trial (bottom). Phonetic boundary  are marked by dotted lines in enrollment and test utterances. ``Er. Seg'' and ``Te. Seg'' show the segments within each phoneme in enrollment and test utterance. The results shown are obtained using the model trained under System (10) in Table~\ref{abalation}.}
\label{local}
\end{figure*}
In this section, we discuss the local interpretation, focusing on trial-level explanations provided by the proposed model.  Fig.~\ref{local} offers a detailed visualization of the model's decision-making process for both non-target and target trials. 
It illustrates the spectrograms of phonetic segments from enrollment and test utterances, alongside their pairwise comparisons, which are represented by the original cosine similarity, the phonetic similarity ($s^i$) scaled to the range $[-\infty, +\infty]$, and the corresponding phonetic weight ($w^i$).

For the non-target trial, the phoneme ``S'', ``N'', ``AH'' and ``AE'' demonstrates a relatively low cosine similarity, leading to a low negative phonetic score, indicating a strong dissimilarity in speaker-specific information between the corresponding segments in the enrollment and test utterances.  Notably, ``N'' is assigned the highest phonetic weight of 1.00, making it a influential factor in determining the final score (-0.975). In contrast, the phoneme ``F'' exhibits a higher phonetic score of 1.41, but its impact on the final decision is significantly diminished due to its much lower phonetic weight of 0.01.

The target trial achieves a higher final score of 0.997. In this case, the phoneme ``N'' demonstrates a high similarity score with phonetic score as 5.44, supported by a substantial phonetic weight 1.00, strongly driving the decision towards a positive match. In contrast, some phonemes, such as ``HH'', exhibit a much lower similarity score of -2.22, but it contribute almost nothing to the final outcome due to their negligible weight.

This detailed analysis provides a fine-grained understanding of the model’s decision-making process at the phonetic level. By emphasizing the interaction between phonetic similarities and their corresponding weights, the figure illustrates how individual phonetic segments collectively influence the speaker verification outcomes. Additionally, with the inclusion of phoneme boundaries, users can further explore the model's behavior by checking or listening to specific segments within a particular phoneme of interest, enabling deeper insights into the relationship between acoustic features and model predictions.

\subsection{Global Interpretation}
\begin{figure*}[htbp]
\centering
\includegraphics[width=18cm]{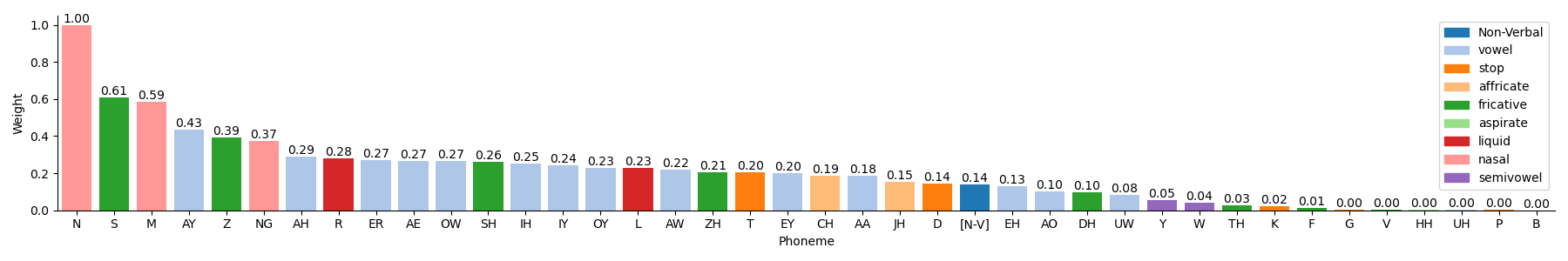}
\caption{Phonetic weight distribution across different phonemes. The results shown are obtained using the model trained under System (10) in Table~\ref{abalation}.}
\label{global}
\end{figure*}

The phonetic weights reflect the overall decision logic of the model by indicating how much each phonetic score contributes to the final score. In Fig.~\ref{global}, we present the rank of weights for each phoneme. Phonemes with higher weights, such as ``N", ``S", and ``M" significantly influence the model's predictions in the decision-making process. Conversely, phonemes like ``B", ``P", and ``HH" receive minimal weight, suggesting they have a negligible impact on the model's outcomes. The weight distribution reveals that certain phonemes play a more prominent role in shaping the prediction score, offering valuable insights into the model's decision-making criteria. 

It is worth mentioning that this global interpretation aligns with human perception in FSC~\cite{moez2016phonetic,1162664,amino2009speaker}, as nasals and vowels (e.g., ``N", ``M", and ``AY") are widely recognized as more discriminative for speaker verification tasks.
Additionally, the discriminative power of fricatives remains a subject of debate~\cite{1162860,kavanagh2012new,moez2016phonetic}. Our results reflect this observation: fricatives such as ``S" and ``Z" show strong discriminative power, while other fricatives, like ``HH" and ``F", receive lower weights, indicating limited influence.

\begin{table}[htbp]
\centering
\caption{Speaker verification performance between models trained on samples of different durations. {The systems are trained under System (4), (6), (10), (12) and (14) in Table~\ref{abalation}}, and evaluated on Vox1-O with the whole test sample (``Whole Utt"), the first 5 seconds (``5s"), and the first 2 seconds (``2s"). “Mix” indicates that the training duration  in each mini-batch is randomly selected from 1 to 5 seconds}
\label{dur-performance}
\begin{tabular}{ccccccc}
\toprule
\multirow{2}{*}{Dur} & \multicolumn{2}{c}{Whole Utt.} & \multicolumn{2}{c}{5s} & \multicolumn{2}{c}{2s} \\  
\cmidrule(lr){2-3} \cmidrule(lr){4-5} \cmidrule(lr){6-7} 
& EER & minDCF & EER & minDCF & EER & minDCF \\ \midrule
1s & 8.411 & 0.556 & 8.818 & 0.594& 12.547 & 0.811 \\
2s & 5.946 & 0.447 & 6.204 & 0.501 & 10.973 & 0.750 \\
3s & 5.673 & 0.439 & 6.053 & 0.479 & 12.047 & 0.790 \\
4s & 6.364 & 0.440 & 6.622 & 0.516 & 15.216 & 0.851 \\
5s & 6.563&  0.457& 7.029& 0.514 &19.248 &0.857  \\ \hdashline 
Mix & 6.869 & 0.491 & 7.379 &0.556 & 12.190 & 0.815\\
\bottomrule
\end{tabular}
\end{table}

\begin{figure*}[!t]
\centering
\includegraphics[width=17cm]{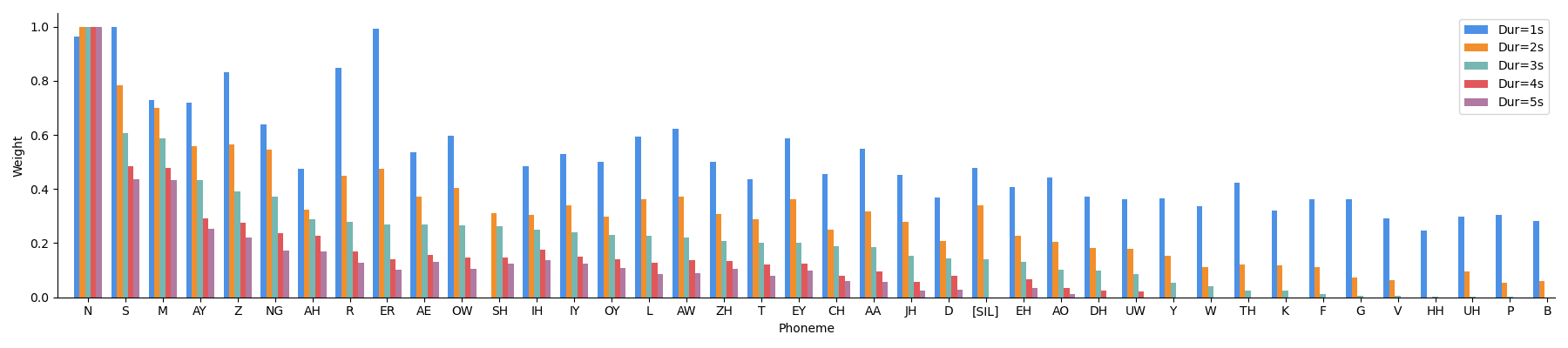}
\caption{The phonetic weight for models trained with various duration. The configuration of these models are same with System (4), (6), (10), (12) and (14) in Table~\ref{abalation} respectively.}
\label{dur}
\end{figure*}
\subsection{Impact of Sample Durations}


As a crucial hyperparameter, the duration of training samples significantly impacts the model's performance especially in short-duration test conditions. As shown in Table~\ref{dur-performance}, while models trained with sample durations ranging from 2 to 5 seconds exhibit comparable performance on the full-duration test set, a notable performance drop occurs for the model trained with 5-second samples when evaluated on a shortened 2-second test set. This drop is commonly attributed to the model's limited generalization ability, caused by the mismatch between training samples and test samples. Using our PhiNET, we provide a deep explanations about this mismatch by analyzing the model's potential cognitive biases. 

Fig.~\ref{dur} visualizes the phonetic weights of models trained with different duration. Across all configurations, the ranking of phonetic weights remains relatively similar, with phonemes like ``N", ``S", and ``M" consistently contributing the most. However, with longer training durations {(e.g. with 5 seconds), the model relies heavily on a few dominant phonemes to make decisions.} 
In contrast, models trained on shorter samples, e.g. with 1 second, 
exhibit a more evenly distributed weight pattern. This broader distribution suggests that model relies on a wider range of phonemes to compensate for the limited number of phonemes present in shorter segments.

This observation suggests that the duration plays a role in terms of phonetic contributions to the speaker verification decisions. Enforcing equal attention to all phonemes may lead to a loss of the model’s capacity to capture speaker-relevant information embedded in more informative phonemes. Conversely, excessive dependence on phonetic features can undermine the model's robustness in short-duration scenarios, limiting its stability and generalization.

We further investigated the effect of training with samples of mixed durations. Specifically, for each mini-batch, the duration of the training utterances was randomly selected from 1 to 5 seconds. Under this training scheme, the results show that the phonetic weights converge to 0.5 for almost all phonemes, indicating that the model fails to learn meaningful distinctions among phonemes for speaker recognition. We hypothesize that each phoneme may have a distinct optimal weight for different training durations. When the duration varies across training samples, the model is exposed to implicit label inconsistency, which confuses the learning process and prevents it from capturing phoneme-specific discriminative patterns. As a result, its performance on the full test utterances and the 5-second segments is close to the worst case, as shown in Table~\ref{dur-performance}. When evaluated with 2-second test samples, the mixed-duration model performs better than models trained with fixed durations of 4 or 5 seconds, but still falls short of the best result achieved with 2-second training samples. This suggests that when the test utterance contains a rich variety of phonemes, the ability to model phoneme-specific discriminative patterns is crucial for achieving high accuracy. In contrast, for short test utterances, the importance of this ability diminishes as the phonemic diversity decreases.

\subsection{Interpretability Evaluation}
\begin{figure*}[htbp]
\centering
\includegraphics[width=18cm]{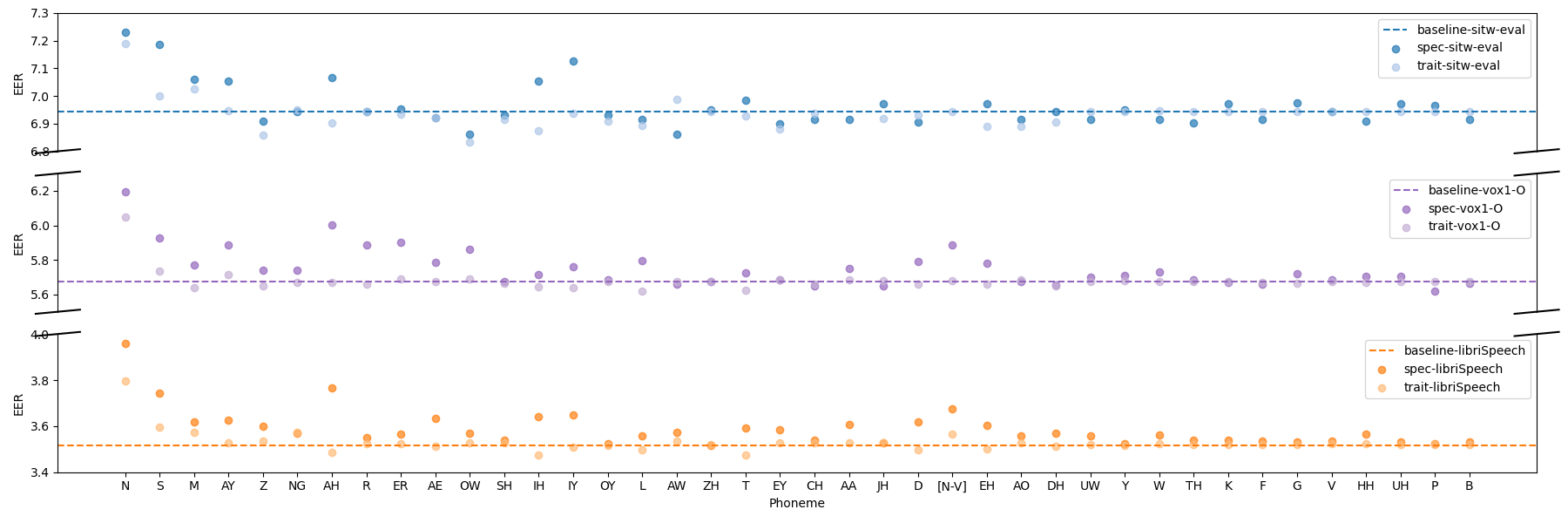}
\caption{Results of {leave-$i$th-phoneme-out} experiments on  SITW-eval, Vox1-O and LibriSpeech. Leave-$i$th-phoneme-out experiments conducted in input spectrogram are shown as  ``spec-sitw-eval'', ``spec-vox1-O'' and ``spec-libriSpeech''. Correspondingly, ``trait'' means the phonetic trait of each phoneme is left out and `baseline' shows the EER of the network without leaving  anything out. }
\label{eval}
\end{figure*}
\subsubsection{Variation of phonetic evidence} As introduced in Section~\ref{eval metric}, we perform leave-$i$th-phoneme-out experiments to evaluate the  performance of global interpretability.
Experiments are conducted on multiple test sets and reported in Fig.~\ref{eval}.  Despite fluctuations in individual test sets, the overall trend is consistently observed across all test sets. 

As expected, performance drops noticeably when high-weighted phonemes are removed. This further confirms that phonemes with higher weights play a more crucial role in the final decision. The findings also show that the  proposed global interpretation method aligns well with human decision process.

\subsubsection{Faithfulness of phonetic trait}
The results in Fig.~\ref{eval} reveal a notable gap between the performance impact of trait removal and spectrogram segment removal.
We hypothesize that this gap arises from the expanded receptive field in the convolutional-based phonetic trait extractor. The stacked convolutional layers increase the receptive field beyond the given phonetic boundaries, capturing adjacent frames. This expansion likely causes the extracted traits to encode speaker information across multiple phonemes, compromising the faithfulness of the system's local interpretability.
Nevertheless, the overall trend across multiple test sets remains consistent and closely aligns with the trend observed from spectrogram segment removal. This indicates that although convolutional networks expand the receptive field, the relative alignment between the output features and the spectrogram is preserved, allowing the model to retain local interpretability.

Table~\ref{abalation} presents the fidelity scores for systems trained under different configurations, {as defined in Section~\ref{eval metric} as the mean absolute difference between the performance changes caused by these two types of removals across all phonemes. }  These results suggest that even when the model architecture remains unchanged, adjustments to the training configuration can enhance faithfulness. A more detailed analysis of this observation will be provided in the next section.

\subsubsection{Example visualization} 
\begin{figure}[h]
    \centering
    \includegraphics[width=0.48\linewidth]{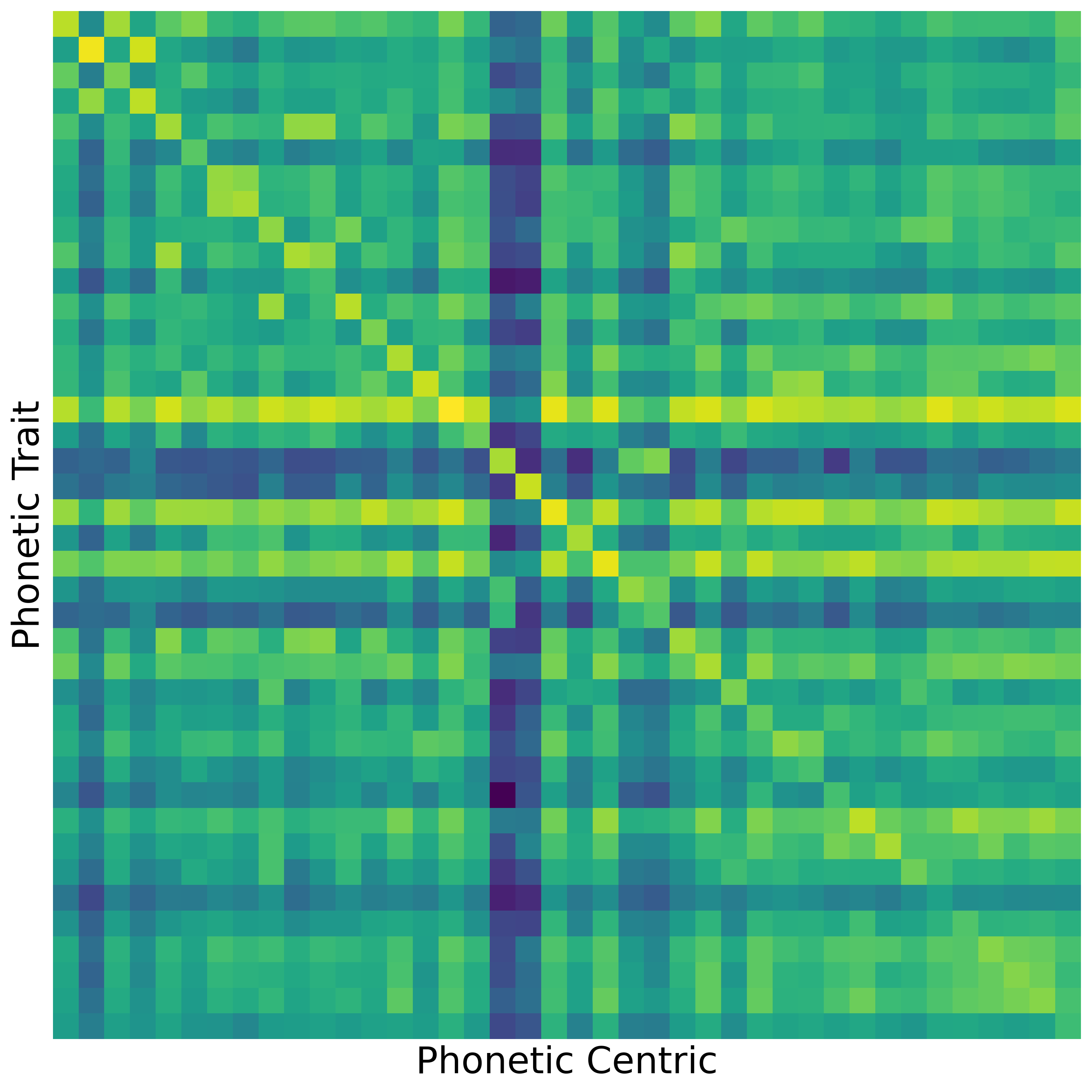} 
    \hfill
    \includegraphics[width=0.48\linewidth]{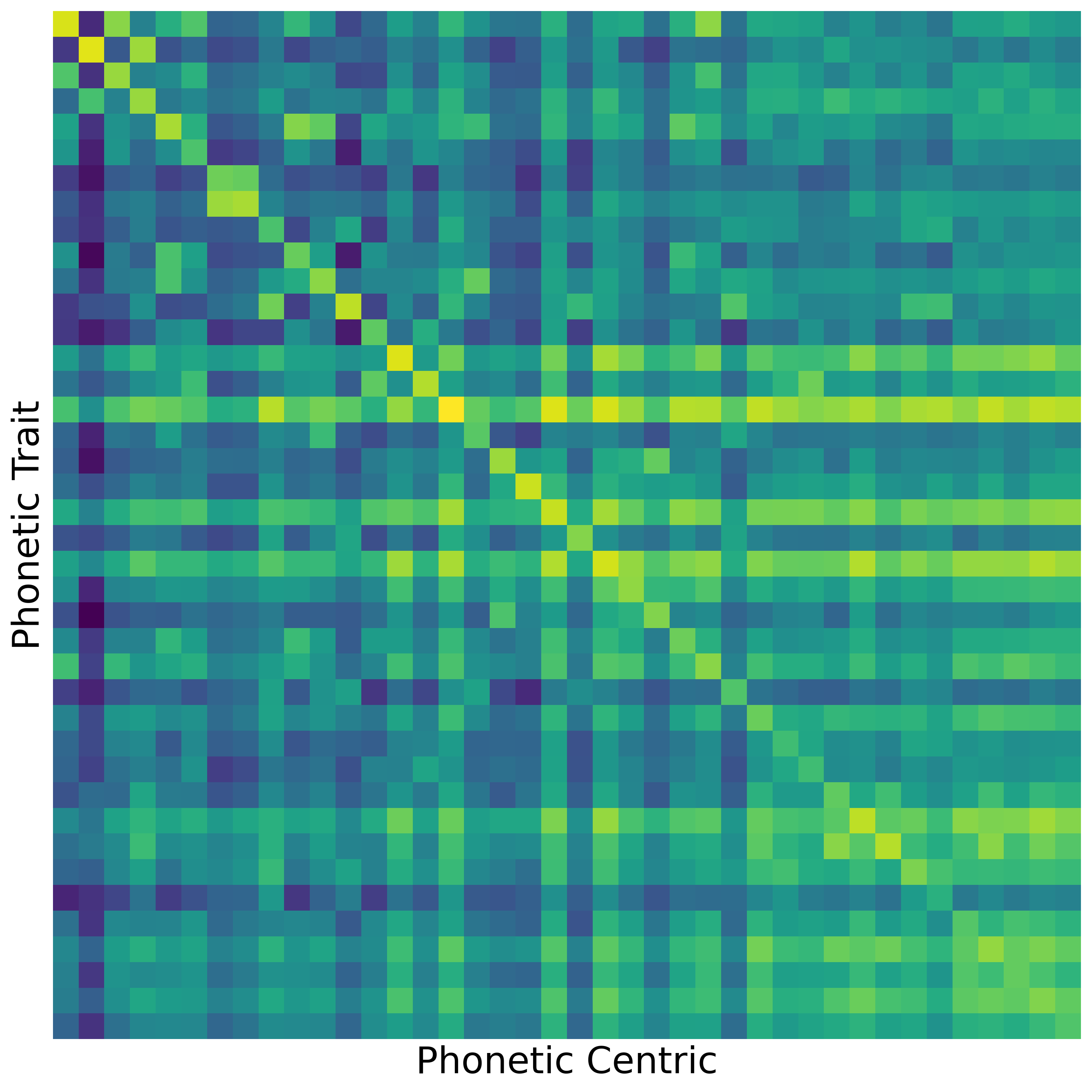} 
    \caption{Similarity heatmaps between the individual phonetic traits and the  centric of each phonetic traits for two speakers in the test set. The results are obtained using the model trained under System (10) in Table~\ref{abalation}. The phoneme order in the heatmaps (top-to-bottom and left-to-right) follows the same order as in Fig~\ref{global}. Brighter points indicate higher cosine similarity.}
    \label{heatmap}
\end{figure}

To provide a more intuitive view of how phonetic traits are either distinct or entangled, we present the similarity heatmaps in Fig.~\ref{heatmap}. For each speaker in the test set of VoxCeleb1, we first compute the centroid of the $i$-th phonetic trait as the average of all $i$-th traits extracted from that speaker’s utterances. We then calculate the average cosine similarity between each phonetic trait and these centroids. The higher cosine similarity indicates the phonetic trait is closer to the centric. We observe that the diagonal elements in Fig.~\ref{heatmap} are noticeably brighter than others, indicating that each phonetic trait is closer to its own centroid than to those of other phonemes. At the same time, we notice that in the lower-right region of the heatmap, the diagonal values tend to be less contrasted with their neighboring entries. This suggests that the traits corresponding to these phonemes are more easily entangled with others, which is consistent with our previous conclusion that these phonemes are less discriminative for speaker verification.

\subsection{Ablation Study}
\begin{table}[htbp]
\center
\caption{Performance in a comparative study between hyper-parameters used in network training. {The fidelity score quantifies the alignment between phonetic trait-based and spectrogram-based leave-one-out setting. Lower values indicate better faithfulness. } All the networks are trained on VoxCeleb1. ``Dur" represents the duration of seconds of training sample. ``$\alpha$" and ``$\beta$" represent the weights assigned to different components in the loss function.   The bolded values indicate the optimal configurations.}
\label{abalation}
\setlength{\tabcolsep}{0.9mm}
\renewcommand{\arraystretch}{1.2}
\begin{tabular}{c|c|cccccc}
\toprule
\multirow{2}{*}{Model} & \multirow{2}{*}{Idx} & \multirow{2}{*}{Dur} & \multirow{2}{*}{$\alpha$} & \multirow{2}{*}{$\beta$} & \multicolumn{3}{c}{Vox1-O} \\ \cline{6-8} 
 &  &&    && EER($\%$) & minDCF & Fidelity$\downarrow$ \\ \hline
\multicolumn{1}{c|}{ECAPA} &(1) &2s & - & -   &  3.403 & 0.272 &-\\ 
-TDNN&(2)&3s&-&- & 3.541&0.279&-\\
\hline
\multicolumn{1}{c|}{\multirow{11}{*}{PhiNet}} & (3)& 1s & 0 & 0  &  8.366 &0.549&0.048\\
\multicolumn{1}{c|}{} & (4) & 1s & 0.001 & 0.0015   & 8.411 &0.556&0.053\\ 
\multicolumn{1}{c|}{} & (5) &2s & 0 & 0  & 6.244 &0.442&0.068\\
\multicolumn{1}{c|}{} &(6) & 2s & 0.001 & 0.0015  & 5.946 & 0.447&0.073\\
\multicolumn{1}{c|}{} & (7) &3s & 0 & 0 &  5.920 & 0.436&0.110\\
\multicolumn{1}{c|}{} & (8) &3s & 0.001 & 0.0005    & 5.746 &0.439&0.068 \\
\multicolumn{1}{c|}{} & (9) &3s & 0.001 & 0.001  & 5.885 &0.426&0.097 \\
\multicolumn{1}{c|}{} & \textbf{(10)} &\textbf{3s} & \textbf{0.001} & \textbf{0.0015} &  \textbf{5.673} & \textbf{0.439}&\textbf{0.081}\\

\multicolumn{1}{c|}{} & (11) &4s & 0 & 0 & 6.290 & 0.456&0.134\\
\multicolumn{1}{c|}{} & (12) &4s & 0.001 & 0.0015   & 6.364 & 0.440&0.100\\

\multicolumn{1}{c|}{} & (13) &5s & 0 & 0 &   6.574 &0.465&0.133\\
\multicolumn{1}{c|}{} & (14) &5s & 0.001 &  0.0015 &   6.563&0.457&0.113\\
\bottomrule
\end{tabular}

\end{table}
In this section, we compare the system trained on VoxCeleb1 with different configuration. The performance of speaker verification and interpretability is evaluated on Vox1-O. For comparison, we also show the performance of ``black-box'' model ECAPA-TDNN in System (1) and (2) in  Table~\ref{abalation}. Since the setting of 2-second training sample is commonly used for black-box speaker model training~\cite{wang2022wespeaker}, we present results for ECAPA-TDNN trained with 2 and 3 seconds, respectively, for a fair comparison.
\subsubsection{Duration of training samples}
To investigate the impact of training sample duration on system, we compare System (4), (6), (10), (12), and (14) in Table~\ref{abalation}, where the sample duration increases progressively from 1 to 5 seconds. It can be observed that the system achieves optimal speaker verification performance with 3-second samples (System (10)), with both the EER and minDCF reaching their lowest values. Beyond 3 seconds, performance begins to degrade as the duration continues to increase. 
Meanwhile, as the duration increases, the interpretability performance, measured by the Fidelity score, shows a continuous upward trend, indicating that the model becomes less faithful. The Fidelity score increases from 0.053 for 1-second samples (System (4)) to 0.113 for 5-second samples (System (14)). This suggests that longer training sample durations result in weaker alignment between the phonetic traits and their corresponding spectrogram regions, thereby reducing the faithfulness of the model's local interpretability.
The speaker verification performance and Fidelity scores indicates a clear trade-off: longer durations improve speaker verification accuracy up to a point but compromise local interpretability.
The similar trend is observed for systems trained exclusively with the loss of $\mathcal{L}_{\text{veri}}$ (Systems (3), (5), (7), (11), and (13)), where performance improves up to 3 seconds before deteriorating with longer durations. Meanwhile, Fidelity score increases progressively from 0.048 to 0.133.
We hypothesize that this behavior is due to the varying amount of phonetic information present in samples of different durations. Shorter samples may lack sufficient phonetic diversity to reliably distinguish between speakers, but the limited number of phonemes allows the system to explore each phoneme thoroughly, which benefits interpretability. Conversely, longer training samples may enhance the phonetic information encoded in the phonetic traits due to expanding receptive field, enabling the model to capture more general speaker characteristics within these traits.

\subsubsection{Impact of $\mathcal{L}_{\text{pho}}$}
In this section, we analyze the impact of incorporating $\mathcal{L}_{\text{pho}}$ into the training process. Models trained solely with $\mathcal{L}_{\text{veri}}$ correspond to Systems (3), (5), (7), (11), and (13) in Table~\ref{abalation}, each evaluated with different training sample durations. In contrast, Systems (4), (6), (10), (12), and (14) integrate $\mathcal{L}_{\text{pho}}$ alongside $\mathcal{L}_{\text{veri}}$ using the same weighting parameters $\alpha$ and $\beta$. Regarding interpretability, as measured by the Fidelity score, the effect of $\mathcal{L}_{\text{pho}}$ remains inconsistent. For training durations of 1 and 2 seconds, models incorporating $\mathcal{L}_{\text{pho}}$ show a slight increase in Fidelity scores. However, for longer durations of 3, 4, and 5 seconds, Fidelity scores tend to decrease when $\mathcal{L}_{\text{pho}}$ is applied. 

\begin{table}[htbp]
\center
\caption{ Impact of different the weight of $\alpha$ and $\beta$ on model performance during training.}
\label{split-weight}
\renewcommand{\arraystretch}{1.2}
\begin{tabular}{cccccc}
\toprule
 \multirow{2}{*}{Dur} & \multirow{2}{*}{$\alpha$} & \multirow{2}{*}{$\beta$}& \multicolumn{3}{c}{Vox1-O} \\ \cline{4-6} 
&  &  &   EER($\%$) & minDCF & Fidelity$\downarrow$ \\ \hline
\multirow{3}{*}{3s}  &0.001 & 0.0015 & 5.673  &0.439 &0.081 \\
&0.0012 & 0.0017 & 5.689 &0.435 & 0.084\\ 
&0.0015& 0.002 &  6.244& 0.459& 0.076\\\hdashline
\multirow{3}{*}{5s}  &0.001 & 0.0015 & 6.563 & 0.457& 0.113\\
&0.0012 & 0.0017 & 6.465 &0.462 &0.129 \\
&0.0015 & 0.002 & 6.385 &0.451 &0.113 \\
\bottomrule
\end{tabular}

\end{table}

Since $\mathcal{L}_{\text{veri}}$ may converge faster for longer training samples because
the model is easier to distinguish between speakers in longer utterances, we further explore the effect of increasing $\alpha$ and $\beta$ for training durations of 5 seconds. The corresponding results compared with configuration of 3 seconds are presented in Table~\ref{split-weight}. These findings suggest that while the weighting of $\mathcal{L}_{\text{pho}}$ requires careful tuning for different configurations, its incorporation has the potential to enhance overall system performance by leveraging phonetic information more effectively.

\subsubsection{Weights in loss function}
We finetune the weight in $\mathcal{L}_{\text{pho}}$ using a fixed $\alpha$ and adjust $\beta$ in Systems (8), (9), and (10) of Table~\ref{abalation}. The EER, minDCF, and Fidelity score for these three systems are similar. However, we observed that the system becomes prone to crashing during training when we try a bigger $\beta$. Despite this, we believe that balancing inter-speaker differences and intra-speaker consistency within the same phonemes is a worthwhile direction and plan to explore it further in future work. In the following sections, unless otherwise specified, we assume that $\alpha$ is 0.001 and $\beta$ is 0.0015 by default.

\subsubsection{Dimension of mapping vector}

\begin{table}[htbp]
\center
\caption{Performance evaluation of models trained with different dimension of mapping vector. $c$ is the dimension
of the mapping vector in eq.~\ref{map2}. }
\label{dimension}
\renewcommand{\arraystretch}{1.2}
\begin{tabular}{cccc}
\toprule
 \multirow{2}{*}{$c$} & \multicolumn{3}{c}{Vox1-O} \\ \cline{2-4} 
& EER($\%$) & minDCF & Fidelity$\downarrow$ \\ \hline
 {1}& {5.808} & {0.448}&{0.108}\\
2& 5.673 &0.439 &0.081\\
3& 5.585 &0.417 &0.095\\
4&5.955& 0.448&0.057\\

\bottomrule
\end{tabular}

\end{table}

We map the raw cosine similarity of each pair of phonetic traits to the range of $[-\infty, +\infty]$ in Eq. (\ref{map2}).  To finetune the dimension of the mapping vector, we set the dimension $c$ to values from 1 to 4, as shown in Table~\ref{dimension}.
The results indicate that speaker verification performance initially improves with increasing $c$, but begins to degrade when $c$ is set to 4. In terms of Fidelity score, we observe a decreasing trend as $c$ increases. Balancing both performance and faithfulness, we select $c$ as 2 for our subsequent experiments.
\begin{table}[htbp]
\center
\caption{ Impact of data augmentation on model performance during training.}
\label{augmentation}
\renewcommand{\arraystretch}{1.2}
\begin{tabular}{cccccc}
\toprule
 \multirow{2}{*}{Loss} & \multirow{2}{*}{$K$} & \multirow{2}{*}{Aug}& \multicolumn{3}{c}{Vox1-O} \\ \cline{4-6} 
&  &  &   EER($\%$) & minDCF & Fidelity$\downarrow$ \\ \hline
\multirow{6}{*}{$\mathcal{L}_{veri}$}  &32 & \checkmark &  6.100 & 0.445&0.083\\
&32 & \texttimes &  6.571 & 0.483&0.100\\ 
&64& \checkmark & 5.840 & 0.430&0.096\\
 &64& \texttimes & 6.747 & 0.486&0.147\\
 & 128&\checkmark & 5.920 & 0.436&0.110\\
 & 128&\texttimes  &7.393  & 0.528&0.141\\\hdashline
\multirow{2}{*}{$\mathcal{L}_{all}$}  &128 & \checkmark & 5.673 & 0.439&0.081\\
&128 & \texttimes & 7.409 & 0.526&0.140\\
\bottomrule
\end{tabular}

\end{table}
\subsubsection{Augmentation during training}
We conducted experiments to evaluate the impact of data augmentation for different training configuration. The systems trained with and without augmentation methods such as adding noise and reverberation are shown in Table~\ref{augmentation}. The results reveal that the absence of augmentation leads to a noticeable decline in both speaker verification performance and faithfullness.
This outcome aligns with our expectations, highlighting the crucial role of data augmentation in enhancing the robustness of the model. Moreover, it underscores that augmentation not only benefits verification accuracy but also plays a key role in maintaining the interpretability of our proposed speaker verification system.

\begin{table}[htbp]
\center
\caption{Performance evaluation of models trained with different function to generate phonetic weights.}
\label{weight-generate}
\renewcommand{\arraystretch}{1.2}
\begin{tabular}{cccc}
\toprule
 \multirow{2}{*}{Function} & \multicolumn{3}{c}{Vox1-O} \\ \cline{2-4} 
& EER($\%$) & minDCF & Fidelity$\downarrow$ \\ \hline
 {Min-Max}& {5.673} & {0.439}&{0.081}\\
Sigmoid& 5.981 &0.444 &0.092\\
Min-Shift& 5.705 &0.433 &0.097\\
ReLU&5.967& 0.458&0.096\\

\bottomrule
\end{tabular}

\end{table}

\subsubsection{Generation of phonetic importance weight}
In our settings, we use a trainable vector, normalized by min-max scaling, as the weight for each phonetic trait in Eq. (\ref{weight}). To explore alternative methods, we replaced min-max normalization with other functions that map the trainable vector to non-negative values. These functions include the Sigmoid function, min-shift normalization, and the ReLU function. 
For the Sigmoid and ReLU functions, we apply the standard Sigmoid and ReLU operations directly on $\bm{\hat{w}}$. For min-shift normalization, $\bm{w}$ is generated by subtracting the minimum value from $\bm{\hat{w}}$ as $\bm{w}=\bm{\hat{w}}-min(\bm{\hat{w}})$. The results are shown in Table~\ref{weight-generate}. 

An interesting observation from these systems is that when the weights are constrained strictly within the range of 0 to 1, weight generated with min-max scaling outperforms Sigmoid function. While when the weights are only constrained to be non-negative, min-shift function surpasses ReLU function. We assume that both min-max and min-shift produce weights that take into account the relationships between all phonetic weights. In contrast, the Sigmoid and ReLU functions generate outputs based solely on individual phonetic weightings.
This observation leads us to hypothesize that speaker verification is a dynamic process in which the importance of a specific phoneme is context-dependent and dynamically related to the weighting of other phonemes.
\subsection{Comparative Study Across Multiple Datasets}
\begin{table*}[htbp]
    \centering
    \caption{Performance of networks evaluated on the Vox1-O/E/H, SITW and LibriSpeech. The system is trained on training set of VoxCeleb1 (`Vox1') and training set of VoxCeleb2 (``Vox2''). ``Pre-trained" refers to the pre-trained ECAPA-TDNN baseline released from WeSpeaker, while ``Reproduced" denotes the ECAPA-TDNN trained with same strategy as our model.}
    \label{various}
    \setlength{\tabcolsep}{0.8mm}
    \renewcommand{\arraystretch}{1}

    \begin{tabular}{cccccccccccccc}
        \toprule
        \multirow{2}{*}{System}&\multirow{2}{*}{Dur}& \multirow{2}{*}{Params} &
         \multirow{2}{*}{Training} &
        \multicolumn{2}{c}{Vox1-O} &
        \multicolumn{2}{c}{Vox1-E} & 
        \multicolumn{2}{c}{Vox1-H} & 
        \multicolumn{2}{c}{SITW-eval} & 
        \multicolumn{2}{c}{LibriSpeech} \\
        \cmidrule(lr){5-6} \cmidrule(lr){7-8} \cmidrule(lr){9-10} \cmidrule(lr){11-12} \cmidrule(lr){13-14} 
       && (Million) &Data& EER & Fidelity & EER & Fidelity& EER & Fidelity& EER& Fidelity& EER & Fidelity\\
        \midrule
        \makecell{ECAPA-TDNN \\ (Pre-trained)}&2s &6.19 & Vox2 & 1.223& - & 1.387 & - & 2.567 & - 
        & 2.352  & - &0.696 &- \\ \hdashline
        \multirow{3}{*}{\makecell{ECAPA-TDNN \\ (Reproduced)}}&2s &\multirow{3}{*}{5.40} & Vox1 & 3.403& - & - & - & - & - 
        & 4.893  & - & 2.061&-  \\
        &3s&&Vox1&3.541 & - & - & - & -  & - 
        & 4.547 & - &2.280 &-  \\
        &2s&&Vox2&2.648&-&2.794&-& 5.474&-&3.318&-&1.685&-\\ \hline
        \multirow{2}{*}{PhiNet}&\multirow{2}{*}{3s}&\multirow{2}{*}{4.81}&Vox1& 5.673 & 0.081& - & - & - & - 
        & 6.944 &0.079  &   3.518&0.060\\ 
         & &&Vox2 &4.186&0.094 &4.398& 0.076&8.839 &0.137&4.921& 0.110& 3.130&0.066
         \\ \hline
         
        \bottomrule
     
    \end{tabular}
\end{table*}
We present the performance of models trained on larger datasets of Vox2 in Table~\ref{various}. 
As expected, speaker verification performance improves with more training samples. However, when evaluating interpretability using fidelity score, we observe that an increased number of training samples does not yield benefits to faithfulness.
Especially for difficult test sets that may contain background noise, such as ``Vox1-H" and ``SITW-eval", the faithfulness of the model degrades significantly as the training dataset grows.
We hypothesize two potential reasons for this behavior. First, it may be attributed to the robustness of the phoneme boundaries provided by the phoneme recognizer. For challenging trials, the phoneme recognizer struggles to provide accurate phoneme boundaries. Second, as the receptive field of the phonetic trait extractor expands, phonetic trait main contains information extracted from adjacent  frames.
With the training set increase, the model becomes more effective at extracting speaker-discriminative features from adjacent  frames. Consequently, when the phonetic traits of a specific phoneme are not particularly discriminative, the model may rely on features from adjacent frames, which could reduce the overall faithfulness. We believe this phenomenon may also reflect a tradeoff between interpretability and high performance for our model.


Compared to the black-box model, our model has a slightly smaller parameter size due to the absence of the utterance-level layer. We also observe that the performance of the released pre-trained ECAPA-TDNN is better than that of the reproduced model trained under our pipeline, indicating that the training mechanism (including augmentation design, optimization method, and pooling strategy) has a non-negligible impact on the final performance. While our model’s EER still exhibits a performance gap relative to the SOTA black-box model, the gap remains within a reasonable and acceptable range. Importantly, across multiple evaluation sets, our model consistently demonstrates strong generalization capability.

We believe this performance difference does not solely arise from the model architecture itself, but also reflects the fact that interpretable speaker verification models are still in the early stage of development, where dedicated training mechanisms tailored for interpretability have not yet been systematically explored. As a result, one promising future direction is to investigate training strategies that are specifically designed for interpretable models, rather than directly inheriting training mechanisms developed for black-box models.

As the first study of self-interpretable speaker verification, we acknowledge that several challenges remain. For instance, phoneme may not represent the optimal minimal unit in the context of speaker verification. 
Considering the characteristic of this model, we believe that future research should focus on disentangling speaker identity from spoken content to further improve model performance.
Another promising direction is to explore whether PhiNet can be adopted as a fine-tuning step on top of strong black-box models (e.g., ECAPA-TDNN). Such a hybrid strategy may leverage the performance of state-of-the-art models while preserving the interpretability offered by our framework, which we leave for future investigation.

\section{Conclusion}
\label{conclusion}

We study speaker verification with phonetic interpretability while maintaining high accuracy. 
The technique allows users to automate the process of fine-grained voice comparison,  visualize the phonetic trait, and identify each phoneme's contribution to the final decision.
Experiment results show that our system achieves comparable accuracy while significantly improving interpretability.
These findings suggest that phoneme-based analysis can be a promising direction for improving the transparency of speaker verification systems.

This work lays the groundwork for developing self-interpretable speaker verification networks. We demonstrate the feasibility of generating meaningful explanations for speaker verification decisions. Despite the challenges, this work marks a significant step toward building systems that are not only accurate but also trustworthy and suitable for high-accountability applications.

\bibliography{ref.bib}
\bibliographystyle{IEEEtran}


 




\vfill

\end{document}